\documentclass[aps,prl,superscriptaddress,amsmath,twocolumn,10pt]{revtex4}

\usepackage{color,hangcaption,hhline,psfrag,rotating,amssymb}
\usepackage[hang,nooneline]{subfigure}
\usepackage{dcolumn}
\usepackage{verbatim}
\usepackage{bm}

\begin{document}
\title{Prediction of the $D_s^*$ width from a calculation of its radiative decay in full lattice QCD}

\author{G.~C.~Donald}
\email[]{donaldg@tcd.ie}
\affiliation{School of Mathematics, Trinity College, Dublin 2, Ireland}
\author{C.~T.~H.~Davies}
\email[]{christine.davies@glasgow.ac.uk}
\affiliation{SUPA, School of Physics and Astronomy, University of Glasgow, Glasgow, G12 8QQ, UK}
\author{J.~Koponen}
\affiliation{SUPA, School of Physics and Astronomy, University of Glasgow, Glasgow, G12 8QQ, UK}
\author{G. P. Lepage}
\affiliation{Laboratory of Elementary-Particle Physics, Cornell University, Ithaca, New York 14853, USA}

\collaboration{HPQCD collaboration}
\homepage{http://www.physics.gla.ac.uk/HPQCD}
\noaffiliation

\date{\today}

\begin{abstract}
We determine the rate for $D_s^* \rightarrow D_s \gamma$ for the 
first time from lattice QCD and include the full effect of 
$u$, $d$ and $s$ sea quarks. The valence quarks are implemented 
using the Highly Improved Staggered Quark (HISQ) formalism 
and we normalise the vector current nonperturbatively. 
We obtain $M(D_s^*)-M(D_s)$ of 148(4) MeV, in good agreement with 
experiment. The value of the decay constant of the $D_s^*$ we 
find to be 274(6) MeV, so that $f_{D_s^*}/f_{D_s} = 1.10(2)$. 
For the radiative decay we find $\Gamma(D_s^* \rightarrow D_s \gamma)$ = 0.066(26) keV. 
Given the experimental branching fraction for this decay 
we predict a total width for the $D_s^*$ of 0.070(28) keV, 
making this the longest lived charged vector meson. 
\end{abstract}


\maketitle

\section{Introduction.} 
\label{sec:intro}

Lattice QCD calculations offer the prospect of increasingly 
accurate results for both hadron masses and simple hadronic 
parameters, decay constants and form factors, 
that give information about the hadron's internal 
structure. 
The hadronic parameters can be directly connected 
to experiment since they control the 
 rate for simple weak and electromagnetic processes.  
For weak processes the connection to experiment involves 
an element from the Cabibbo-Kobayashi-Maskawa (CKM) matrix 
and we can use this to determine the CKM elements. 
For electromagnetic processes there is no CKM factor 
and so the comparison is more 
direct. 
Where electromagnetic decay rates have been measured 
experimentally, they then provide strong tests of 
the calculation of QCD matrix elements that are 
very similar to the ones that appear in weak decays. 

For the electromagnetic decays of 
charged charmed vector mesons we have a particularly interesting 
situation 
in which there is significant destructive interference 
between radiation of the photon from the $c$ quark 
and from the light quark. This makes the results very 
sensitive to the different contributions to the 
decay matrix element and therefore a stringent test 
of the calculation. 

Here we study $D_s^* \rightarrow D_s \gamma $ 
decay using lattice 
QCD for the first time. We are able to calculate the 
rate for the decay accurately 
by using gluon field 
configurations that include the full effect of 
$u$, $d$ and $s$ quarks in the sea at multiple 
values of the lattice spacing, by having a 
formalism for the quarks with very small 
discretisation errors and because we are able
to normalise the current that couples to the 
photon fully nonperturbatively.  
As described below, we find that the effective form factor 
for the $D_s^*$ decay is only 
20\% of that for photon emission from the $s$ quark 
alone, so that the total rate for the electromagnetic 
decay is very highly suppressed. 
It nevertheless represents 94.2(7)\% of the 
branching fraction~\cite{pdg} and so we find that
the total width of the $D_s^*$
is the narrowest of any of the 
vector mesons containing a $c$ quark. Only the $B_s^*$ is 
expected to be narrower.   

We also provide further important tests of 
QCD through
our accurate determination of the $D_s^*$ mass 
and its decay constant. The mass determination 
adds to the growing set of gold-plated meson masses
from lattice QCD that are tested at 
the few MeV level against experiment. 
The decay constant allows us to determine the 
leptonic annihilation rate for the $D_s^*$, which 
is much larger than that of the $D_s$ because it 
is not helicity suppressed. The rate is (only) 5
orders of magnitude smaller than the electromagnetic 
decay rate. Although small, this is the largest 
branching fraction for annihilation to a $W$ 
boson for any vector meson.

\section{Lattice Calculation}
\label{sec:lattice}

For the lattice QCD calculation we use the Highly Improved Staggered 
Quark action~\cite{HISQ_PRD} for all the valence quarks. This action 
has very small discretisation errors, making it an excellent action 
for both $c$ and $s$~\cite{HISQ_PRD, HISQ_PRL, Dsdecayconst, jpsi}.  
We calculate HISQ propagators on gluon field configurations generated 
by the MILC collaboration that include $u$, $d$ and $s$ sea quarks 
using the asqtad formalism~\cite{MILCconfigs}. Table~\ref{tab:params} 
gives the parameters of the ensembles of configurations we use, 
with two different 
lattice spacing values and two different $u/d$ sea quark masses.  

\begin{table}
\begin{tabular}{lllllll}
\hline
\hline
Set &  $r_1/a$ & $au_0m_{l}^{asq}$ & $au_0m_{s}^{asq}$ & $L_s/a \times L_t/a$ & $n_{cfg}$ & T/a\\
\hline
1 &  2.647(3) & 0.005 & 0.05 & 24 $\times$ 64 & 2088 & 12,15,18 \\
2 &  2.618(3) & 0.01 & 0.05 & 20 $\times$ 64 & 2259 & 12,15,18 \\
\hline
3 & 3.699(3) & 0.0062 & 0.031 & 28 $\times$ 96 & 1911 & 16,20,23 \\
\hline 
\hline
\end{tabular}
\caption{Ensembles (sets) of MILC configurations used here. 
Sea 
(asqtad) quark masses $m_{\ell}^{asq}$ ($\ell = u/d$) and $m_s^{asq}$ 
use the MILC convention where $u_0$ is the plaquette 
tadpole parameter. 
The lattice spacing is given in units of $r_1$ after `smoothing'
~\cite{MILCconfigs}. We use $r_1=0.3133(23)$ fm~\cite{Davies:2009tsa}. 
Sets 1 and 2 are `coarse' ($a \approx 0.12$ fm) and set 3, 
`fine' ($a \approx 0.09$ fm).  The lattice size 
is given by $L_s^3 \times L_t$. 
We use 4 time sources on each of the $n_{cfg}$ configurations and 3 values 
of $T/a$ (Fig.~\ref{fig:3ptdiag}). 
}
\label{tab:params}
\end{table}

To tune the $s$ and $c$ quark masses to 
their correct physical values we use the pseudoscalar $\eta_s$ 
and $\eta_c$ meson masses~\cite{Dsdecayconst, dsphi}.
The $\eta_s$ is a fictitious $s\overline{s}$ pseudoscalar 
that is not allowed to decay in lattice QCD, and so 
does not correspond to a particle in the real world.
It is useful because its mass can be accurately determined in 
lattice QCD as  
0.6858(40) GeV~\cite{Davies:2009tsa}. 
In tuning the $c$ quark mass here we must use the 
value of the $\eta_c$ mass~\cite{Dsdecayconst} in a world without 
electromagnetism or $c$ quarks in the sea. We 
take this to be $M_{\eta_c}$=2.985(3) GeV~\cite{Gregory:2010gm}. 

The HISQ $s$ and $c$ 
quark propagators calculated on these gluon fields 
are combined to 
make meson correlators for 
$D_s^*$ and $D_s$ mesons. For the $D_s$ we use the local 
pseudoscalar operator which, in combination with the quark 
mass, is absolutely normalised~\cite{Dsdecayconst}. For the 
$D_s^*$ we use the local vector operator, whose normalisation 
can be determined fully nonperturbatively as described in~\cite{dsphi}. 
The $Z$ factors from~\cite{dsphi} are reproduced in Table~\ref{tab:res}. 
The correlators are fit to a multi-exponential form using a Bayesian 
approach~\cite{gplbayes} so that we can include systematic errors from 
the presence of higher excitations in the correlator when extracting 
the ground-state mass and amplitude. 
The excited state parameters are loosely constrained by prior values and 
widths; splittings between excited states are given 
a prior of 600(300) MeV and amplitudes are 
given priors of 0.01(1.0). 
The ground-state masses obtained from a 6-exponential fit 
are given in Table~\ref{tab:res} along 
with the decay constants determined from the ground-state 
amplitude as discussed in~\cite{Dsdecayconst, jpsi}.

\begin{table*} 
\begin{center} 
\begin{tabular}{cccccccccccc} 
\hline 
\hline 
Set & $am_s$ & $am_c$ & $aM_{D_s}$ & $aM_{D_s^*}$ & $af_{D_s}$ & $af_{D_s^*}/Z_{cs}$ & $V_c(0)/Z_{cc}$ & $V_s(0)/Z_{ss}$ & $Z_{cs}$ & $Z_{cc}$ & $Z_{ss}$\\ 
\hline 
1 & 0.0489 & 0.622 & 1.18976(17) & 1.2800(7) & 0.15435(18) & 0.1813(10) & 1.21(9) & 3.02(15) & 1.027(3) & 0.9896(11) & 1.007(12) \\ 
\hline 
2 & 0.0496 & 0.630 & 1.20209(21) & 1.2942(9) & 0.15641(24) & 0.1793(11) & 1.33(6) & 3.24(12) & 1.020(10) & 0.9894(8) & 1.003(9) \\ 
\hline 
3 & 0.0337 & 0.413 & 0.84701(12) & 0.9112(5) & 0.10790(11) & 0.1202(5) &  1.22(7) & 2.95(18) & 1.009(2) & 1.0049(10) & 1.009(11)\\ 
\hline 
\hline 
\end{tabular} 
\caption{Results for the masses of the $D_s$ and $D_s^*$ mesons and the $D_s$ and $D_s^*$ decay constants in lattice units for the HISQ valence $c$ and $s$ masses given in columns 2 ad 3. Columns 8 and 9 give the vector form factors at $q^2 =0$ for the cases where the photon couples to the $c$ or $s$ quarks. We also give the $Z$ factors we need to include for the $\bar{c}s$, $\bar{c}c$ and $\bar{s}s$ vector currents.
}
\label{tab:res} 
\end{center} 
\end{table*}

\begin{figure}
\centering
\includegraphics[angle=-90,width=0.45\textwidth]{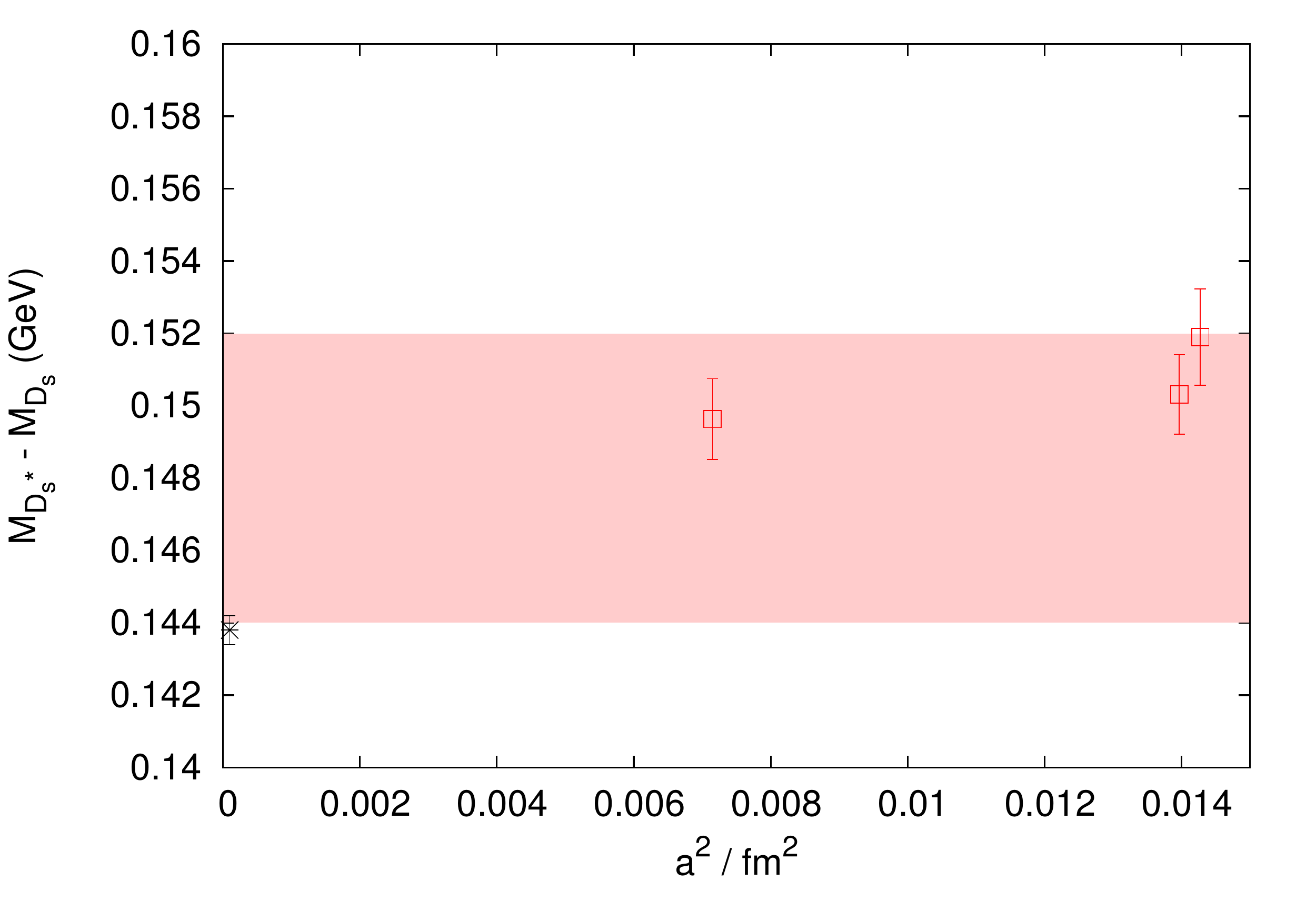}
\includegraphics[angle=-90,width=0.45\textwidth]{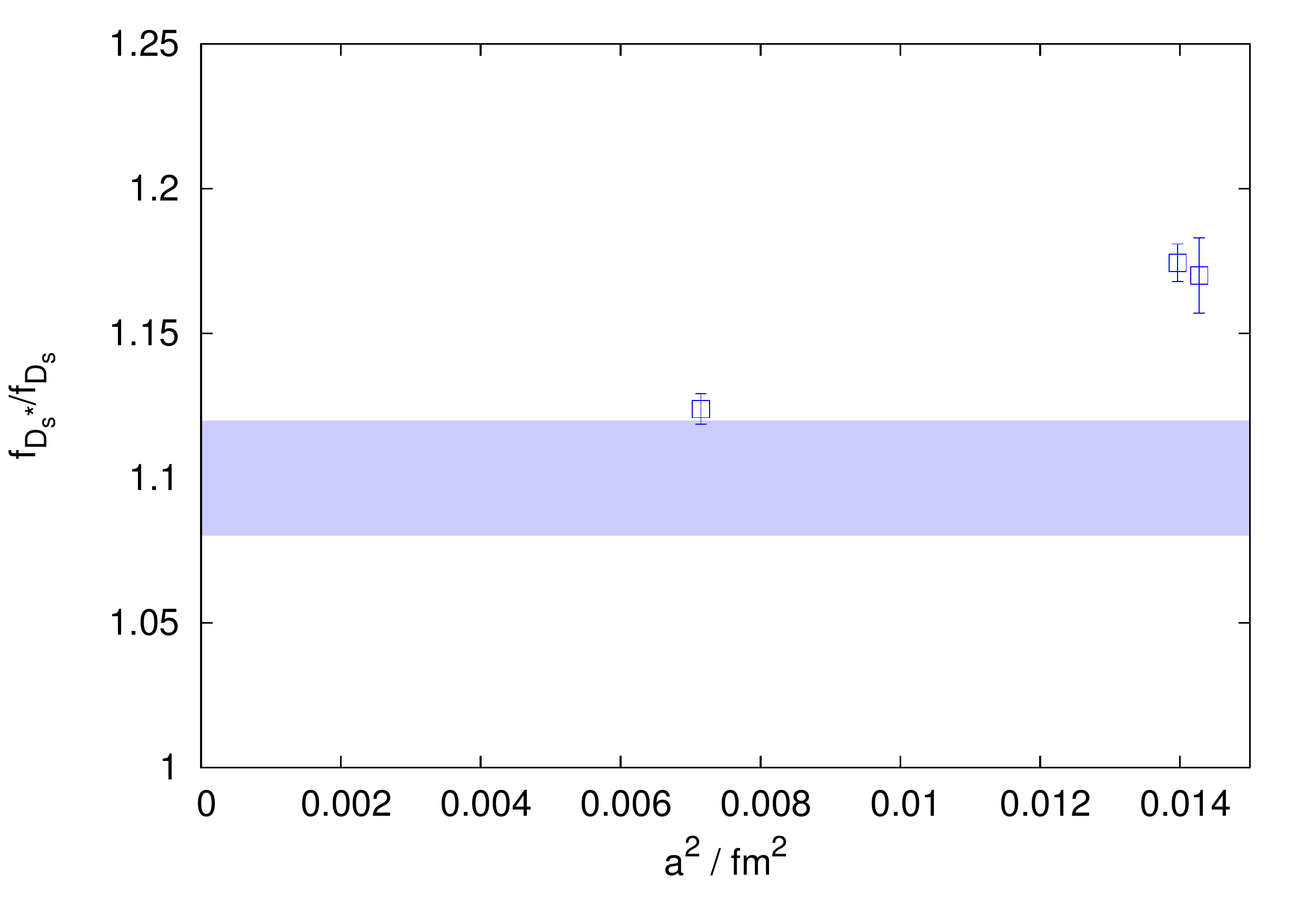}
\caption{
Upper plot: The difference in mass between the $c\overline{s}$ 
vector $D_s^*$ and pseudoscalar $D_s$ as a function of $a^2$. 
The experimental result~\cite{pdg} is plotted with a black burst. 
The red shaded band gives the lattice 
result with its total error 
in the continuum and chiral limits from a simple fit described 
in the text.  \\
Lower plot: A similar plot for the ratio of $D_s^*$ and $D_s$ 
decay constants. The blue shaded band shows the physical result obtained 
from the corresponding fit.
}
\label{fig:mf}
\end{figure}

The mass difference between $D_s^*$ and $D_s$ and decay 
constant ratio are plotted as a function of $a$ in Fig.~\ref{fig:mf}. 
We see mild dependence on the lattice spacing, and almost none on 
the sea quark mass. 
We fit the lattice results as a function of $a$ and sea quark mass to
\begin{equation} 
f(a^2,\delta x_m) = f_0 \times \left( 1 + \sum_{i=1}^5 c_i \left(\frac{am_c}{2}\right)^{2i} + \sum_{j=1}^2 \chi_j (\delta x_m)^j \right)
\label{eq:extrapolate}
\end{equation} 
where $\delta x_m$ is the 
discrepancy between the sea 
quark masses, $2m_l+m_s$, and their physical value in units of the 
physical $s$ quark mass taken from~\cite{Dsdecayconst}. 
We take priors on $c_j$ and $\chi_j$ to be 0.0(1.0) apart from 
$c_1$ which we take to be 0.0(0.3) since $a^2$ errors are 
suppressed by $\alpha_s$ for the HISQ action~\cite{HISQ_PRD}. 

The physical result ($f_0$) that we obtain for $M_{D_s^*}-M_{D_s}$ 
is $148(3)(2)$ MeV, where the first error is from the extrapolation and 
the second from the (correlated) uncertainty in the lattice 
spacing determination~\cite{Davies:2009tsa} which has a 
double impact on hyperfine splittings 
because of their sensitivity to tuning the quark masses~\cite{jpsi}. 
This is in good agreement with the experimental average of 
143.8(4) MeV~\cite{pdg}. Note that there are no additional systematic 
uncertainties from missing electromagnetism or charm-in-the-sea because 
the leading effects of these, already small, cancel between between $D_s^*$ 
and $D_s$~\cite{Dsdecayconst, jpsi}.  

Our physical result for $f_{D_s^*}/f_{D_s}$ is 1.10(2), clearly greater than 
1. Combined with our earlier result for $f_{D_s}$ of 
248.0(2.5) MeV~\cite{Dsdecayconst} (with which we agree here but with 
larger errors) we predict   
$f_{D_s^*} = 274(6)$ MeV.

\begin{figure}
\centering
\includegraphics[width=0.45\textwidth]{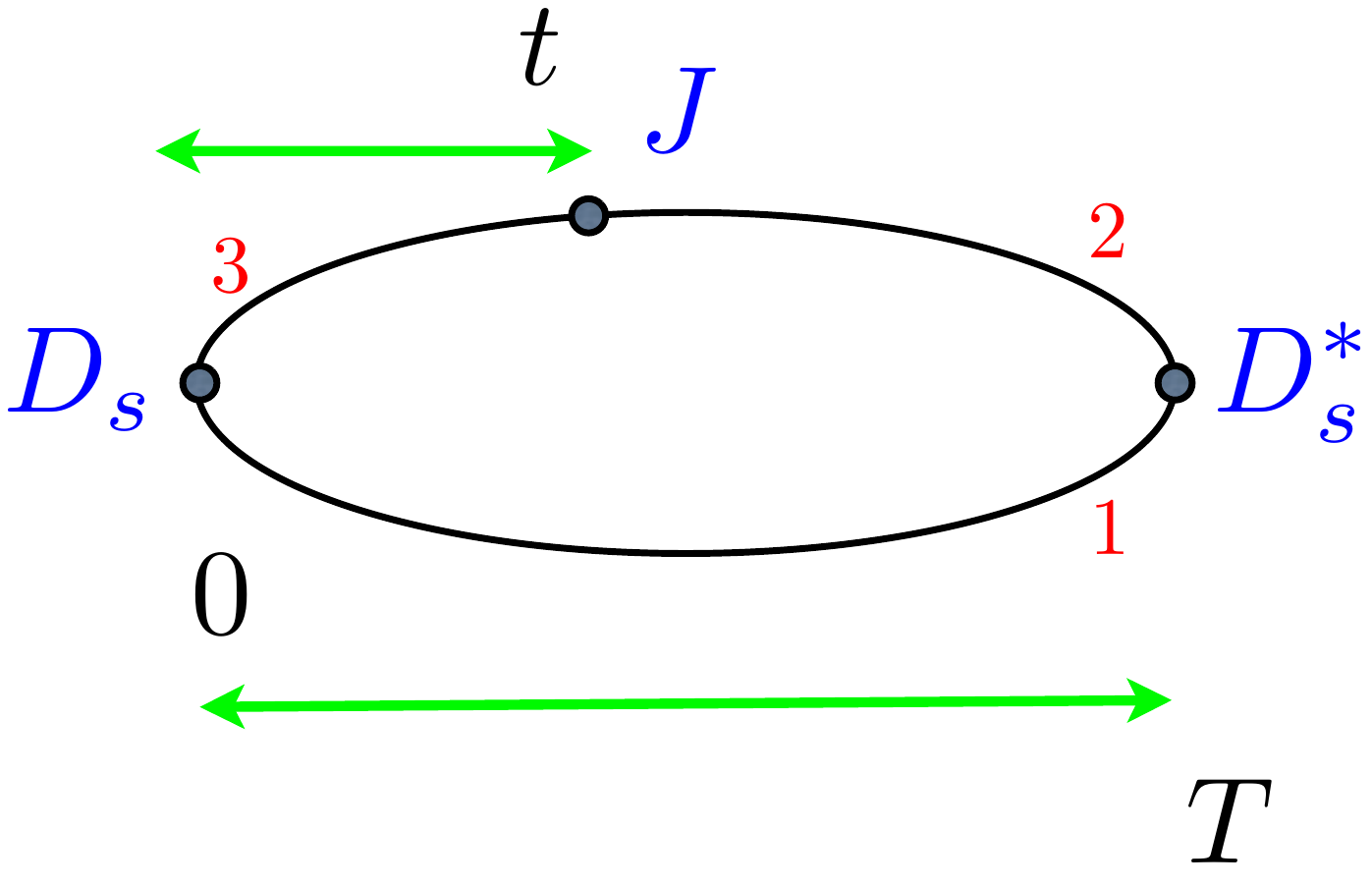}
\caption{A schematic diagram of the 3-point function 
for $D_s^* \rightarrow D_s \gamma$ decay. $J$ is a vector current which 
can couple either to the $s$ quark or the $c$ quark in the $D_s^*$. 
}
\label{fig:3ptdiag}
\end{figure}

\begin{figure}
\centering
\includegraphics[angle=-90,width=0.45\textwidth]{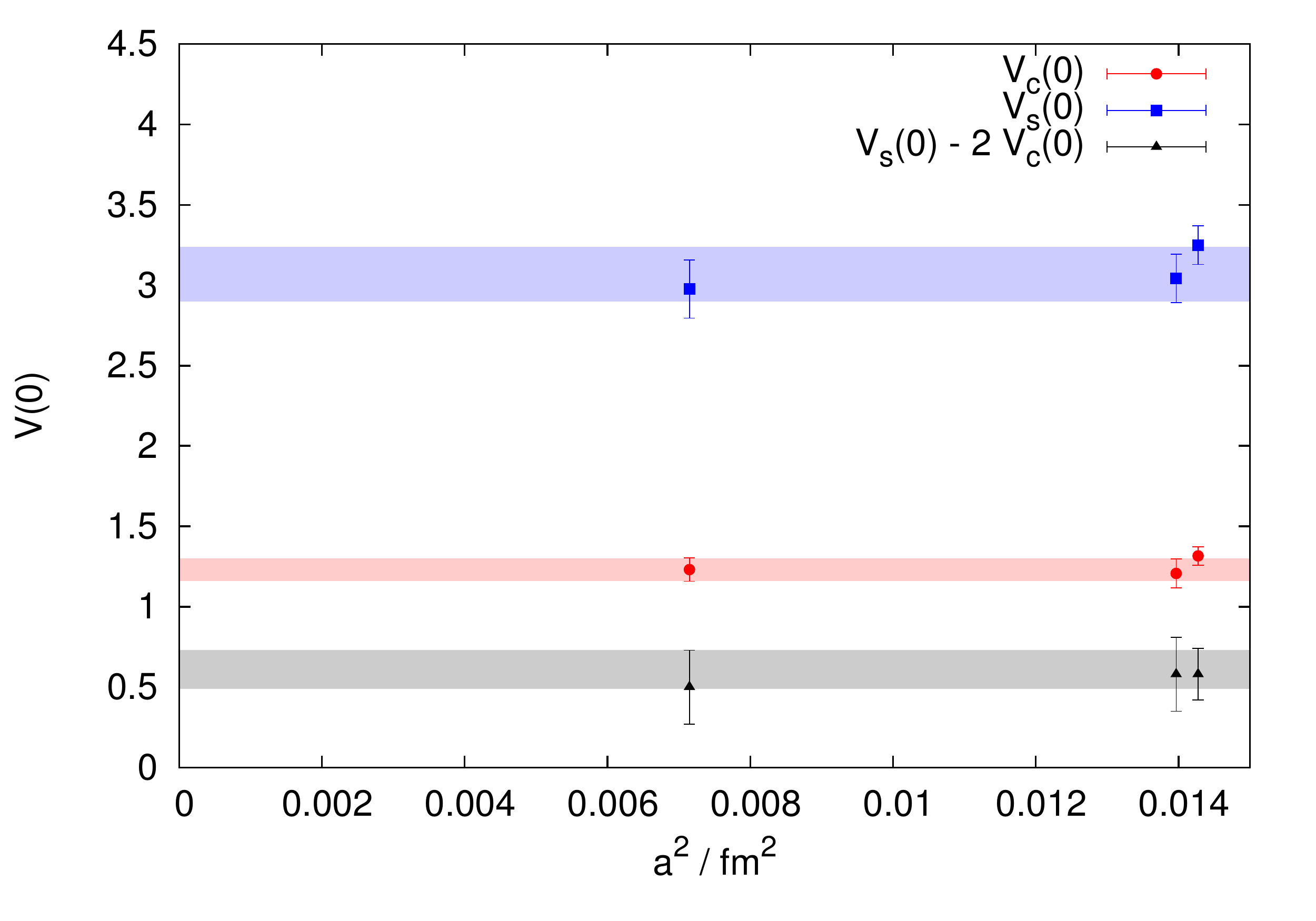}
\caption{
The vector form factor for $D_s^* \rightarrow D_s \gamma$ 
at $q^2=0$ for a transition via a $c\overline{c}$ current 
(denoted $V_c(0)$) and via an $s\overline{s}$ current 
(denoted $V_s(0)$). The form factors are plotted against 
$a^2$. We also show $V_s(0)-2V_c(0)$ which is 
the effective form factor which appears in the rate for 
$D_s^* \rightarrow D_s \gamma$.  
}
\label{fig:v0}
\end{figure}

A 3-point correlation function that allows 
us to calculate the $D_s^*$ to $D_s$ transition matrix 
element is sketched in Fig.~\ref{fig:3ptdiag}, with $J$ 
representing the vector current that couples to the photon.  
We must consider two cases for photon emission at $J$. 
In one, propagators 2 and 3 
are charm quarks and 1 is a strange quark. In case 2, propagators 
2 and 3 are strange quarks and 1 is a charm quark.  
Since the photon produced in $D_s^* \rightarrow D_s \gamma$ 
decay is real, we need to tune the momentum of 
the $D_s$ in the rest frame of the $D_s^*$ so that the 
square of the 4-momentum transferred, $q^2$, is zero. 
This is done by calculating propagator 3 
with a `twisted boundary condition'~\cite{firsttwist, etmctwist}, 
to give it a small, tuned spatial momentum.

When making correlation functions with staggered quarks we 
have a choice of operators because every meson 
comes in 16 `tastes'that differ by effects proportional to $a^2$~\cite{HISQ_PRD}. 
In a 3-point function the taste combinations at the 3 points must cancel. 
Here, for $D_s^* \rightarrow D_s\gamma$, we follow the procedure 
developed for $J/\psi \rightarrow \eta_c \gamma$~\cite{jpsi}. We take 
the $D_s$ to be the `Goldstone' pseudoscalar 
(in taste-spin notation $\gamma_5\otimes\gamma_5$), the $D_s^*$
uses a 1-link operator ($\gamma_0\gamma_{i}\otimes\gamma_0\gamma_{i}\gamma_{j}$) 
and then the vector current is a local vector $\gamma_k\otimes\gamma_k$. 
We can normalise this vector current fully nonperturbatively using the 
techniques described in~\cite{dsphi,jpsi}. 

The 3-point functions for $D_s^* \rightarrow D_s$ are
calculated for all $t$ values from 
$0$ to $T$ and for 3 values of $T$ (see Table~\ref{tab:params})
so that the dependence 
of the function on $t$ and $T$ can be fully mapped 
out. Both 3-point functions are fit simultaneously 
with the 2-point functions for the $D_s$ and $D_s^*$ using 
the operators discussed above at source and sink. 
The fit functions have a multi-exponential form as given in ~\cite{jpsi}, 
and we use the same Bayesian approach and priors described for the 
2-point correlators above.

The quantity that we extract from the fit is the vector 
current matrix 
element $\langle D_s^* | \mathcal{V} | D_s \rangle$
between the ground-state particles in the $D_s$ and $D_s^*$ channels. 
This is related to the 
vector form factor $V(q^2)$ by 
\begin{equation} Z{\langle D_s^*(p^{\prime},\varepsilon) | \mathcal{V}^{\mu} | D_s(p) \rangle} = \frac{2\epsilon^{\mu\alpha\beta t}}{m_{D_s}+m_{D_s^*}} \varepsilon^{*}_{\alpha} p_{\beta} p'_t V(q^2), \label{eq:vff} \end{equation}
where we have allowed for a renormalisation of the lattice vector current. 
Note that for a non-zero answer all the vectors have to point in different directions. 
The $D_s^*$ is at rest so its momentum only has a component in 
the $t$ direction. 
We have two results for the vector form factor, that in which emission is 
from the $s$ quark and that in which emission is from the $c$ quark. 
We give the results separately for $V_s(0)/Z_{s\overline{s}}$ and 
$V_c(0)/Z_{c\overline{c}}$ in Table \ref{tab:res} along with the appropriate 
$Z$ factors from~\cite{dsphi}. The $D_s^*$ masses obtained from the fit 
are consistent with those from the local vector operator 
but with larger uncertainties. 

To calculate the rate for $D_s^* \rightarrow D_s \gamma$ decay we 
need to include factors for the quark and antiquark electric charge, 
which have the same sign because the $D_s^*$ is charged. 
However there is a relative minus sign between the two contributions 
because this transition 
requires a spin-flip to convert a vector into a pseudoscalar. 
Thus 
the effective `total' form factor is $V_{\mathrm{eff}}(0)=[V_s(0)-2V_c(0)]/3$ 
and the partial width for the decay is given by
\begin{equation}
\Gamma_{(D_s^* \to D_s \gamma)} = \alpha_{QED} \frac{4|\vec{q}|^3}{3(M_{D_s}+M_{D_s^*})^2} \frac{|V_s(0) - 2V_c(0)|^2}{9} .
\end{equation}
Here $|\vec{q}|$ is the magnitude of the momentum of 
the $D_s$ in the $D_s^*$ rest frame 
and takes value 138.9(6) MeV using the experimental masses for $D_s$ and $D_s^*$~\cite{pdg}. 

Fig.~\ref{fig:v0} shows our results for $V_s(0)$, $V_c(0)$ and $3V_{\mathrm{eff}}(0)$ 
as a function of lattice spacing. Very little dependence on the lattice 
spacing or sea quark masses is seen. We fit the results as a function 
of $a$ and $\delta {x_m}$ to the form given in Eq.~\ref{eq:extrapolate}, 
obtaining physical results $V_{s}(0)$=3.07(17), $V_{c}(0)$=1.23(7) 
and $3V_{\mathrm{eff}}(0)$=0.61(12). Correlations between $V_c(0$ and 
$V_s(0$ reduce the error in $V_{\mathrm{eff}}(0)$. 
Our result for $V_{\mathrm{eff}}(0)$ gives a partial width for the 
transition of 0.066(26) keV. The error is dominated by the lattice 
statistical error as the result of subtracting two form factors 
that almost cancel.  

Fig.~\ref{fig:v0} shows several interesting features. One is the relative size
of $V_c(0)$ and $V_s(0)$. Since the transition is an M1 transition
we expect the form factor to be proportional to the quark velocity in 
the meson. For a charm-strange meson, if each of the quarks has momentum 
of $\mathcal{O}(\Lambda_{QCD})$, the $s$ quark will be relativistic but 
the $c$ quark will have a velocity of $\Lambda_{QCD}/m_c \approx 1/3$.
We can therefore readily explain the factor of around 3 between the 
two form factors. In fact the factor is less than 3 which means that the 
destructive interference between the two form factors is even more severe,
given that the electric charge ratio is 2. 
We can also compare $V_c(0)$ here to that for $J/\psi \rightarrow \eta_c$ 
decay where we obtained a value of 1.90(7)~\cite{jpsi}. Here we have a significantly 
lower result showing that the form factor is pushed down by the presence 
of a lighter (here $s$) spectator quark. This is consistent with the velocity of 
a $c$ quark in $J/\psi$ being higher than in a $D_s^*$.  

\section{Discussion/Conclusions}
\label{sec:discussion}

We find the partial width for electromagnetic decay of the $D_s^*$ 
to be very small, 0.066(26) keV. The branching fraction for this 
decay is measured to be 94.2(7)\%~\cite{babardsstar}, giving a total width for 
the $D_s^*$ of 0.070(28) keV and a lifetime of $9.4(3.8) \times 10^{-18}$s. 
Estimates for the $D_s^*$ radiative decay width using a variety of 
non-lattice techniques give a range of results~\cite{Casalbuoni} ranging 
from 0.06keV to 0.4keV. 
Our calculation shows that the width is at the lower end of this range. 
Only the $B_s^*$ is likely to be longer-lived~\cite{Casalbuoni}. In that 
case the effective form factor is a sum of $V_s(0)$ and $V_b(0)$ but 
the kinematic factors reduce the rate.  

The only other measured decay rate for the $D_s^*$ is that to $D_s\pi^0$, 
a Zweig-suppressed isospin-violating $P$-wave decay. This has 
the remaining 5.8\% branching 
fraction which, given our total width, corresponds to a partial width of 
$4.0\times 10^{-3}$keV. We can parameterise this hadronic decay in terms 
of a $D_s^*D_s\pi$ coupling as  
\begin{equation} \Gamma_{(D_s* \to D_s \pi^0)} = \frac{g^2_{D_s^*D_s\pi}}{24\pi M^2_{D_s^*}} p_\pi^3 \end{equation}
where $p_{\pi}$ is the $\pi$ momentum in the $D_s^*$ rest frame (47.8(3.2) MeV~\cite{pdg}).
This gives a result for $g_{D_s^*D_s\pi}$ of $0.112(11)_{\mathrm{expt}}(24)_{\mathrm{latt}}$, to be compared to 
that from the Zweig-allowed $D^*$ to $D\pi$ transition of 17.9(1.9)~\cite{cleog}. 

For $D^{*+}$ the electromagnetic decay has a much smaller rate than the 
hadronic decay. Its total width~\cite{babardstarwidth} 
and branching fraction~\cite{pdg} 
yield a partial width for the 
electromagnetic decay of 1.3(3) keV. This is 20(6) times larger than the result 
we find for the $D_s^*$ and implies a much weaker cancellation of 
$V_c(0)$ against $V_d(0)$ than that between $V_c(0)$ and $V_s(0)$ 
in the $D_s^*$. Comparison of $V_c(0)$ for $D_s^*$ to that for 
$J/\psi$~\cite{jpsi} 
indicate that $V_c(0)$ for the $D^*$ could be lower still. 
It also seems likely that $V_d(0)$ for $D^{*+}$ will be higher than 
$V_s(0)$ for $D_s^*$ based on similar arguments. 20\% or larger
shifts would be needed in both of the two directions to reach the experimental 
result. A direct calculation in lattice QCD for the $D^{*+}$ can of course now be 
done following the method we have given here for the $D_s^*$.  

Our result for the $D_s^*$ decay constant shows it to be 
10\% higher than that for the $D_s$. The ratio of vector 
to pseudoscalar decay constants is predicted in Heavy 
Quark Effective Theory to be less than 1~\cite{Neubertpr} in the infinite 
heavy quark mass limit, but to be larger than 1 for 
$c$ and $b$ quarks when $1/m_Q$ effects are included. 
A recent value for $f_{D_s^*}/f_{D_s}$ from QCD sum rules 
is 1.32(10)~\cite{Gelhausen:2013wia}, and from lattice 
QCD with $u/d$ quarks (only) in the sea is 
1.26(3)~\cite{Becirevic:2012ti} (statistical errors only).

Our ratio of 1.10(2) shows that the vector to 
pseudoscalar decay constant ratio 
at $c$ is closer to 1 than these earlier expectations, and that therefore the 
ratio at $b$ can be expected to be even smaller.  
Going in the opposite direction to where the `heavy' 
quark is an $s$ quark, results in lattice 
QCD for the $\eta_s$~\cite{Davies:2009tsa} and experiment 
for the $\phi$~\cite{pdg, dsphi} show a consistent picture 
with a larger ratio: 
$f_{\phi}/f_{\eta_s} = 1.26(2)$. 
Our decay constant calculation is also complemented 
by the good agreement with experiment of our accurate result for 
the $D_s^*$ mass, to be expected of lattice QCD 
calculations with $u$, $d$ and $s$ 
sea quarks for a gold-plated particle.  

Our result $f_{D_s^*} = 274(6)$ MeV can be used to determine the 
weak leptonic decay rate from: 
\begin{equation}
\Gamma_{(D_s^* \rightarrow \ell \nu)} = \frac{G_F^2}{12\pi}|V_{cs}|^2f_{D_s^*}^2{M_{D_s^*}^3}(1-\frac{m_{\ell}^2}{M_{D_s^*}^2})^2(1+\frac{m_{\ell}^2}{2M_{D_s^*}^2}).
\end{equation}
Our result for $f_{D_s^*}$ gives a partial width of $2.4(1)\times 10^{-6}$ keV 
for this decay and hence branching fraction, using 
our determination of the total width, of $3.4(1.4) \times 10^{-5}$. 
This offers potentially the best prospect of measuring a weak 
annihilation rate for a vector meson.

{\it Acknowledgements.} We are grateful to the MILC collaboration 
for the use of their 
gauge configurations and to E. Follana for useful discussions. 
We used the Darwin Supercomputer 
as part of STFC's DiRAC facility jointly
funded by STFC, BIS 
and the Universities of Cambridge and Glasgow. 
This work was funded by STFC, the EU (ITN STRONGnet), the Royal Society and the Wolfson Foundation.

\bibliography{ds-star}

\begin{thebibliography}{19}
\expandafter\ifx\csname natexlab\endcsname\relax\def\natexlab#1{#1}\fi
\expandafter\ifx\csname bibnamefont\endcsname\relax
  \def\bibnamefont#1{#1}\fi
\expandafter\ifx\csname bibfnamefont\endcsname\relax
  \def\bibfnamefont#1{#1}\fi
\expandafter\ifx\csname url\endcsname\relax
  \def\url#1{\texttt{#1}}\fi
\expandafter\ifx\csname urlprefix\endcsname\relax\def\urlprefix{URL }\fi
\providecommand*{\bibinfo}[2]{#2}
\providecommand*{\eprint}[1]{#1}
\providecommand*{\url}[1]{#1}
\begingroup\makeatletter
 \@temptokena{%
  \expandafter\ifx\csname citenamefont\endcsname\relax
   \DeclareRobustCommand\citenamefont{\@firstofone}%
   \global\let\citenamefont\citenamefont
   \global\expandafter\let\csname citenamefont \expandafter\endcsname\csname
  citenamefont \endcsname
  \fi
 }\if@filesw\immediate\write\@auxout{\the\@temptokena}\fi
\expandafter\endgroup\the\@temptokena

\bibitem[{Beringer \emph{et~al.}(2012)\citenamefont{Beringer}
  \emph{et~al.}}]{pdg}
\bibinfo{author}{\bibfnamefont{J.}~\bibnamefont{Beringer}} \emph{et~al.}
  (\bibinfo{collaboration}{Particle Data Group}), \bibinfo{journal}{Phys.Rev.}
  \textbf{\bibinfo{volume}{D86}}, \bibinfo{pages}{010001}
  (\bibinfo{year}{2012}).

\bibitem[{\citenamefont{Follana} \emph{et~al.}(2007)\citenamefont{Follana,
  Mason, Davies, Hornbostel, Lepage} \emph{et~al.}}]{HISQ_PRD}
\bibinfo{author}{\bibfnamefont{E.}~\bibnamefont{Follana}},
  \bibinfo{author}{\bibfnamefont{Q.}~\bibnamefont{Mason}},
  \bibinfo{author}{\bibfnamefont{C.}~\bibnamefont{Davies}},
  \bibinfo{author}{\bibfnamefont{K.}~\bibnamefont{Hornbostel}},
  \bibinfo{author}{\bibfnamefont{G.~P.} \bibnamefont{Lepage}}, \emph{et~al.}
  (\bibinfo{collaboration}{HPQCD and UKQCD Collaborations}),
  \bibinfo{journal}{Phys.Rev.} \textbf{\bibinfo{volume}{D75}},
  \bibinfo{pages}{054502} (\bibinfo{year}{2007}), \eprint{hep-lat/0610092}.

\bibitem[{\citenamefont{Follana} \emph{et~al.}(2008)\citenamefont{Follana,
  Davies, Lepage, and Shigemitsu}}]{HISQ_PRL}
\bibinfo{author}{\bibfnamefont{E.}~\bibnamefont{Follana}},
  \bibinfo{author}{\bibfnamefont{C.}~\bibnamefont{Davies}},
  \bibinfo{author}{\bibfnamefont{G.}~\bibnamefont{Lepage}}, \bibnamefont{and}
  \bibinfo{author}{\bibfnamefont{J.}~\bibnamefont{Shigemitsu}}
  (\bibinfo{collaboration}{HPQCD and UKQCD Collaborations}),
  \bibinfo{journal}{Phys.Rev.Lett.} \textbf{\bibinfo{volume}{100}},
  \bibinfo{pages}{062002} (\bibinfo{year}{2008}), \eprint{0706.1726}.

\bibitem[{\citenamefont{Davies}
  \emph{et~al.}(2010{\natexlab{a}})\citenamefont{Davies, McNeile, Follana,
  Lepage, Na} \emph{et~al.}}]{Dsdecayconst}
\bibinfo{author}{\bibfnamefont{C.}~\bibnamefont{Davies}},
  \bibinfo{author}{\bibfnamefont{C.}~\bibnamefont{McNeile}},
  \bibinfo{author}{\bibfnamefont{E.}~\bibnamefont{Follana}},
  \bibinfo{author}{\bibfnamefont{G.}~\bibnamefont{Lepage}},
  \bibinfo{author}{\bibfnamefont{H.}~\bibnamefont{Na}}, \emph{et~al.}
  (\bibinfo{collaboration}{HPQCD Collaboration}), \bibinfo{journal}{Phys.Rev.}
  \textbf{\bibinfo{volume}{D82}}, \bibinfo{pages}{114504}
  (\bibinfo{year}{2010}{\natexlab{a}}), \eprint{1008.4018}.

\bibitem[{\citenamefont{Donald} \emph{et~al.}(2012)\citenamefont{Donald,
  Davies, Dowdall, Follana, Hornbostel} \emph{et~al.}}]{jpsi}
\bibinfo{author}{\bibfnamefont{G.}~\bibnamefont{Donald}},
  \bibinfo{author}{\bibfnamefont{C.}~\bibnamefont{Davies}},
  \bibinfo{author}{\bibfnamefont{R.}~\bibnamefont{Dowdall}},
  \bibinfo{author}{\bibfnamefont{E.}~\bibnamefont{Follana}},
  \bibinfo{author}{\bibfnamefont{K.}~\bibnamefont{Hornbostel}}, \emph{et~al.}
  (\bibinfo{collaboration}{HPQCD Collaboration}), \bibinfo{journal}{Phys.Rev.}
  \textbf{\bibinfo{volume}{D86}}, \bibinfo{pages}{094501}
  (\bibinfo{year}{2012}), \eprint{1208.2855}.

\bibitem[{\citenamefont{Bazavov} \emph{et~al.}(2010)\citenamefont{Bazavov,
  Toussaint, Bernard, Laiho, DeTar} \emph{et~al.}}]{MILCconfigs}
\bibinfo{author}{\bibfnamefont{A.}~\bibnamefont{Bazavov}},
  \bibinfo{author}{\bibfnamefont{D.}~\bibnamefont{Toussaint}},
  \bibinfo{author}{\bibfnamefont{C.}~\bibnamefont{Bernard}},
  \bibinfo{author}{\bibfnamefont{J.}~\bibnamefont{Laiho}},
  \bibinfo{author}{\bibfnamefont{C.}~\bibnamefont{DeTar}}, \emph{et~al.},
  \bibinfo{journal}{Rev.Mod.Phys.} \textbf{\bibinfo{volume}{82}},
  \bibinfo{pages}{1349} (\bibinfo{year}{2010}), \eprint{0903.3598}.

\bibitem[{\citenamefont{Davies}
  \emph{et~al.}(2010{\natexlab{b}})\citenamefont{Davies, Follana, Kendall,
  Lepage, and McNeile}}]{Davies:2009tsa}
\bibinfo{author}{\bibfnamefont{C.}~\bibnamefont{Davies}},
  \bibinfo{author}{\bibfnamefont{E.}~\bibnamefont{Follana}},
  \bibinfo{author}{\bibfnamefont{I.}~\bibnamefont{Kendall}},
  \bibinfo{author}{\bibfnamefont{G.}~\bibnamefont{Lepage}}, \bibnamefont{and}
  \bibinfo{author}{\bibfnamefont{C.}~\bibnamefont{McNeile}}
  (\bibinfo{collaboration}{HPQCD Collaboration}), \bibinfo{journal}{Phys.Rev.}
  \textbf{\bibinfo{volume}{D81}}, \bibinfo{pages}{034506}
  (\bibinfo{year}{2010}{\natexlab{b}}), \eprint{0910.1229}.

\bibitem[{\citenamefont{Donald} \emph{et~al.}(2013)\citenamefont{Donald,
  Davies, Koponen, and Lepage}}]{dsphi}
\bibinfo{author}{\bibfnamefont{G.}~\bibnamefont{Donald}},
  \bibinfo{author}{\bibfnamefont{C.}~\bibnamefont{Davies}},
  \bibinfo{author}{\bibfnamefont{J.}~\bibnamefont{Koponen}}, \bibnamefont{and}
  \bibinfo{author}{\bibfnamefont{G.}~\bibnamefont{Lepage}}
  (\bibinfo{collaboration}{HPQCD Collaboration})  (\bibinfo{year}{2013}),
  \eprint{1311.6669}.

\bibitem[{\citenamefont{Gregory} \emph{et~al.}(2011)\citenamefont{Gregory,
  Davies, Kendall, Koponen, Wong} \emph{et~al.}}]{Gregory:2010gm}
\bibinfo{author}{\bibfnamefont{E.~B.} \bibnamefont{Gregory}},
  \bibinfo{author}{\bibfnamefont{C.~T.~H.} \bibnamefont{Davies}},
  \bibinfo{author}{\bibfnamefont{I.~D.} \bibnamefont{Kendall}},
  \bibinfo{author}{\bibfnamefont{J.}~\bibnamefont{Koponen}},
  \bibinfo{author}{\bibfnamefont{K.}~\bibnamefont{Wong}}, \emph{et~al.}
  (\bibinfo{collaboration}{HPQCD collaboration}), \bibinfo{journal}{Phys. Rev.}
  \textbf{\bibinfo{volume}{D83}}, \bibinfo{pages}{014506}
  (\bibinfo{year}{2011}), \eprint{1010.3848}.

\bibitem[{Lepage \emph{et~al.}(2002)\citenamefont{Lepage}
  \emph{et~al.}}]{gplbayes}
\bibinfo{author}{\bibfnamefont{G.~P.} \bibnamefont{Lepage}} \emph{et~al.},
  \bibinfo{journal}{Nucl. Phys. Proc. Suppl.} \textbf{\bibinfo{volume}{106}},
  \bibinfo{pages}{12} (\bibinfo{year}{2002}), \eprint{hep-lat/0110175}.

\bibitem[{\citenamefont{de~Divitiis}
  \emph{et~al.}(2004)\citenamefont{de~Divitiis, Petronzio, and
  Tantalo}}]{firsttwist}
\bibinfo{author}{\bibfnamefont{G.}~\bibnamefont{de~Divitiis}},
  \bibinfo{author}{\bibfnamefont{R.}~\bibnamefont{Petronzio}},
  \bibnamefont{and} \bibinfo{author}{\bibfnamefont{N.}~\bibnamefont{Tantalo}},
  \bibinfo{journal}{Phys.Lett.} \textbf{\bibinfo{volume}{B595}},
  \bibinfo{pages}{408} (\bibinfo{year}{2004}), \eprint{hep-lat/0405002}.

\bibitem[{\citenamefont{Guadagnoli}
  \emph{et~al.}(2006)\citenamefont{Guadagnoli, Mescia, and Simula}}]{etmctwist}
\bibinfo{author}{\bibfnamefont{D.}~\bibnamefont{Guadagnoli}},
  \bibinfo{author}{\bibfnamefont{F.}~\bibnamefont{Mescia}}, \bibnamefont{and}
  \bibinfo{author}{\bibfnamefont{S.}~\bibnamefont{Simula}},
  \bibinfo{journal}{Phys.Rev.} \textbf{\bibinfo{volume}{D73}},
  \bibinfo{pages}{114504} (\bibinfo{year}{2006}), \eprint{hep-lat/0512020}.

\bibitem[{Aubert \emph{et~al.}(2005)\citenamefont{Aubert}
  \emph{et~al.}}]{babardsstar}
\bibinfo{author}{\bibfnamefont{B.}~\bibnamefont{Aubert}} \emph{et~al.}
  (\bibinfo{collaboration}{BaBar Collaboration}), \bibinfo{journal}{Phys.Rev.}
  \textbf{\bibinfo{volume}{D72}}, \bibinfo{pages}{091101}
  (\bibinfo{year}{2005}), \eprint{hep-ex/0508039}.

\bibitem[{\citenamefont{Casalbuoni}
  \emph{et~al.}(1997)\citenamefont{Casalbuoni, Deandrea, Di~Bartolomeo, Gatto,
  Feruglio} \emph{et~al.}}]{Casalbuoni}
\bibinfo{author}{\bibfnamefont{R.}~\bibnamefont{Casalbuoni}},
  \bibinfo{author}{\bibfnamefont{A.}~\bibnamefont{Deandrea}},
  \bibinfo{author}{\bibfnamefont{N.}~\bibnamefont{Di~Bartolomeo}},
  \bibinfo{author}{\bibfnamefont{R.}~\bibnamefont{Gatto}},
  \bibinfo{author}{\bibfnamefont{F.}~\bibnamefont{Feruglio}}, \emph{et~al.},
  \bibinfo{journal}{Phys.Rept.} \textbf{\bibinfo{volume}{281}},
  \bibinfo{pages}{145} (\bibinfo{year}{1997}), \eprint{hep-ph/9605342}.

\bibitem[{Anastassov \emph{et~al.}(2002)\citenamefont{Anastassov}
  \emph{et~al.}}]{cleog}
\bibinfo{author}{\bibfnamefont{A.}~\bibnamefont{Anastassov}} \emph{et~al.}
  (\bibinfo{collaboration}{CLEO Collaboration}), \bibinfo{journal}{Phys.Rev.}
  \textbf{\bibinfo{volume}{D65}}, \bibinfo{pages}{032003}
  (\bibinfo{year}{2002}), \eprint{hep-ex/0108043}.

\bibitem[{Lees \emph{et~al.}(2013)\citenamefont{Lees}
  \emph{et~al.}}]{babardstarwidth}
\bibinfo{author}{\bibfnamefont{J.}~\bibnamefont{Lees}} \emph{et~al.}
  (\bibinfo{collaboration}{BaBar Collaboration}),
  \bibinfo{journal}{Phys.Rev.Lett.} \textbf{\bibinfo{volume}{111}},
  \bibinfo{pages}{101802} (\bibinfo{year}{2013}), \eprint{1305.1575}.

\bibitem[{\citenamefont{Neubert}(1994)}]{Neubertpr}
\bibinfo{author}{\bibfnamefont{M.}~\bibnamefont{Neubert}},
  \bibinfo{journal}{Phys.Rept.} \textbf{\bibinfo{volume}{245}},
  \bibinfo{pages}{259} (\bibinfo{year}{1994}), \eprint{hep-ph/9306320}.

\bibitem[{\citenamefont{Gelhausen} \emph{et~al.}(2013)\citenamefont{Gelhausen,
  Khodjamirian, Pivovarov, and Rosenthal}}]{Gelhausen:2013wia}
\bibinfo{author}{\bibfnamefont{P.}~\bibnamefont{Gelhausen}},
  \bibinfo{author}{\bibfnamefont{A.}~\bibnamefont{Khodjamirian}},
  \bibinfo{author}{\bibfnamefont{A.~A.} \bibnamefont{Pivovarov}},
  \bibnamefont{and}
  \bibinfo{author}{\bibfnamefont{D.}~\bibnamefont{Rosenthal}},
  \bibinfo{journal}{Phys.Rev.} \textbf{\bibinfo{volume}{D88}},
  \bibinfo{pages}{014015} (\bibinfo{year}{2013}), \eprint{1305.5432}.

\bibitem[{\citenamefont{Becirevic} \emph{et~al.}(2012)\citenamefont{Becirevic,
  Lubicz, Sanfilippo, Simula, and Tarantino}}]{Becirevic:2012ti}
\bibinfo{author}{\bibfnamefont{D.}~\bibnamefont{Becirevic}},
  \bibinfo{author}{\bibfnamefont{V.}~\bibnamefont{Lubicz}},
  \bibinfo{author}{\bibfnamefont{F.}~\bibnamefont{Sanfilippo}},
  \bibinfo{author}{\bibfnamefont{S.}~\bibnamefont{Simula}}, \bibnamefont{and}
  \bibinfo{author}{\bibfnamefont{C.}~\bibnamefont{Tarantino}},
  \bibinfo{journal}{JHEP} \textbf{\bibinfo{volume}{1202}}, \bibinfo{pages}{042}
  (\bibinfo{year}{2012}), \eprint{1201.4039}.

\end{thebibliography}

\end{document}